\documentstyle[epsf,12pt]{article}
\newcommand{\newsection}{    
\setcounter{equation}{0}
\section}
\def\appendix#1{
\setcounter{section}{0}
\addtocounter{section}{1}
\setcounter{equation}{0}
\renewcommand{\thesection}{\Alph{section}}
\section*{Appendix \thesection\protect\indent #1}
\addcontentsline{toc}{section}{Appendix \thesection\ \ \ #1}
}
\newcommand{\rf}[1]{(\ref{#1})}
\newcommand{\eq}[1]{Eq.~(\ref{#1})}

\def\be{\begin{equation}}
\def\ee{\end{equation}}
\newcommand{\beq}{\begin{equation}}
\newcommand{\eeq}{\end{equation}}
\newcommand{\bea}{\begin{eqnarray}}
\newcommand{\eea}{\end{eqnarray}}

\begin{document}
\input{epsf}
\topmargin 0pt
\oddsidemargin 5mm
\headheight 0pt
\headsep 0pt
\topskip 9mm
\pagestyle{empty}

\hfill NBI-HE-98-28

\hfill SPhT-98/119
\addtolength{\baselineskip}{0.20\baselineskip}
\begin{center}
\vspace{26pt}
{\large \bf Hamiltonian Cycles on Random Eulerian Triangulations}


\vspace{26pt}

\vspace{18pt}
{\sl E.\ Guitter}\hspace{0.025cm}\footnote{E-mail:
guitter@spht.saclay.cea.fr} \\
CEA-Saclay, Service de Physique Th\'{e}orique\\
F-91191 Gif-sur-Yvette Cedex, France \\

\vspace{18pt}
{\sl C. Kristjansen}\hspace{0.025cm}\footnote{E-mail: kristjan@alf.nbi.dk}
and {\sl J.L.\ Nielsen}\hspace{0.025cm}\footnote{E-mail:
langgard@alf.nbi.dk}
\\ 
\vspace{6pt}
The Niels Bohr Institute \\
 Blegdamsvej 17,
DK-2100 Copenhagen \O, Denmark \\
\end{center}
\vspace{20pt} 
\begin{center}
Abstract
\end{center}
\noindent
A random Eulerian triangulation is a random triangulation where an
{\it even} number of triangles meet at any given vertex. We argue that
the central charge increases by one if 
the fully packed $O(n)$ model is defined on a random Eulerian triangulation  
instead of an ordinary random triangulation. 
Considering the case $n\rightarrow 0$, this implies that
the system of random Eulerian triangulations equipped with Hamiltonian
cycles describes a $c=-1$  matter field coupled to 2D quantum gravity
as opposed to the system of usual random triangulations equipped with
Hamiltonian cycles which has $c=-2$. Hence, in this case 
one should see a change in the entropy exponent from the value
$\gamma=-1$ to the {\it irrational} value
$\gamma=\frac{-1-\sqrt{13}}{6}=-0.76759...$ when going from a usual random
triangulation to an Eulerian one. A direct enumeration of
configurations confirms this change in $\gamma$.

\vfill{\noindent PACS codes: 05.20.y, 04.60.Nc, 02.10.Eb }
\newpage

\pagestyle{plain}
\setcounter{page}{1}

\newsection{Introduction}
In order to describe the thermodynamic properties of geometrical objects
like polymers and membranes, it is important to be able to enumerate the 
possible configurations of such objects. One class of problems concerns
the {\it folding} of these objects. The simplest non trivial examples are 
the compact folding of a {\it self-avoiding} polymer, in connection with 
protein folding, and the folding onto itself of a {\it phantom} 
(non self-avoiding) polymerised membrane. Traditionally one has studied the
folding statistics of self-avoiding polymers by embedding them on a {\it 
regular} two-dimensional lattice and identifying the possible folded states of
the polymer with the Hamiltonian cycles on the lattice~\cite{polymers}.
A Hamiltonian cycle is a closed curve which visits each vertex of the lattice
once and only once. Similarly, to describe the possible folded states
of a phantom polymerised membrane one can approximate the membrane by a 
regular two-dimensional lattice and identify the possible folded states of 
the membrane with those of the lattice. 
In order to facilitate our discussion, we shall restrict ourselves to
considering three-coordinate or equivalently triangular lattices.
The problem of counting the number of folded states of the regular, 
triangulated lattice has been proven to be equivalent to a certain 
three-colouring problem, namely the problem of colouring the links of 
the lattice with three different colours so that no two links which 
belong to the same triangle carry the same colour~\cite{DG94}. 
This three-colouring problem is a classical mathematical problem which 
was solved by Baxter in 1970~\cite{Bax70}. It can also be described as 
the problem of solving the fully packed $O(n)$ model for $n=2$~\cite{Bax70}.
Similarly, the problem of counting the number of Hamiltonian cycles on a 
regular three-coordinate lattice is equivalent to solving the fully packed 
$O(n)$ model in the limit $n\rightarrow 0$. This model is critical and
describes a conformal field theory with central charge $c=-1$~\cite{BN94,BSY94}.

However, the use of a regular lattice to describe folding problems is
clearly limitative. For instance, most membranes in nature are fluid
rather than polymerised, and their modelling requires a {\it random} lattice
instead.  Similarly, one might speculate if the complex dynamics of 
polymers does not call for a random lattice rather than a regular one. 
Making use of the so-called fully packed $O(n)$ models on a 
random lattice
(cf.\ section~\ref{random}) it is possible to generalise the above
mentioned folding problems to a random lattice. There are, however,
some subtleties involved in this process. This is because the full
packing constraint is very sensitive to the local geometry of the 
lattice.  For instance when one solves the Hamiltonian cycle problem 
on a random three-coordinate lattice one finds that 
$c=-2$~\cite{DK90,EGK98,Hig98}. We will explain the reason for this
difference between the regular and the random case and argue that 
if we restrict the class of random triangulations considered to 
so-called random Eulerian triangulations we will find again $c=-1$.
By random Eulerian triangulations we mean random triangulations where 
each vertex is shared by an even number of triangles. Such triangulations 
also appear when we try to generalise the membrane folding problem to 
a random lattice. If we consider simply the fully packed $O(2)$ model 
on a random lattice we do not get a model describing the folding of a 
random triangulation. We do also not get a model that describes the 
generalisation of Baxter's three-colouring problem to a random lattice. 
In order to have a model which describes the generalised three-colour 
problem one must consider a fully packed $O(2)$ model where the length 
of the loops (appearing in the graphical representation of the model, 
cf.\ section~\ref{regular}) is restricted to being {\it even}~\cite{EK97}. 
Still, on a random lattice the edge-three-colouring problem is not 
equivalent to the folding problem. In order to have a model which describes 
the folding problem one must consider the edge colouring problem on a 
random Eulerian triangulation~\cite{EK97,DFEG98} or equivalently the fully 
packed $O(2)$ model on a random Eulerian triangulation (note that the 
condition of the loops of the $O(n)$ model having even length is 
automatically satisfied on an Eulerian triangulation). 
Eulerian triangulations can also be described as triangulations permitting 
a three-colouring of their vertices~\cite{DFEG98}. The folding problem on 
a random lattice can hence be viewed as a double three-colouring problem.

In section~\ref{regular} and~\ref{random} we review the properties of
the fully packed $O(n)$ model on a regular and on a random lattice 
respectively and argue that on a random lattice we should see a change in 
central charge for the fully packed model if we change the class of 
triangulations considered from ordinary random triangulations to
random Eulerian triangulations. In section~\ref{Eulerian} we
specialise to the case $n\rightarrow 0$ and show that Hamiltonian
cycles on a random Eulerian triangulation should provide us with an
explicit realisation of a random surface model with an irrational
value of the entropy exponent, $\gamma$. Section~\ref{solution}
describes a numerical solution of the Hamiltonian cycle problem on a
random Eulerian triangulation and section~\ref{results} contains the
results -- results which support our
arguments. Finally, in section~\ref{conclusion} we conclude and
comment on the similarity  between our Hamiltonian cycle problem and
another combinatorial problem which has recently attracted a lot of
attention, namely the Meander problem~\cite{Meanders}.

\newsection{Dense versus full packing of the $O(n)$-model on a regular 
lattice \label{regular} }
Let us first consider a regular three-coordinate lattice, the
honeycomb lattice, and let us associate to each vertex, $i$, of the
lattice a $n$-component spin vector $\vec{S}_{i}$ with
$\vec{S}_i^2=n$. The $O(n)$-model partition function is then given
by
\beq
Z(T)=\int \prod_i d\vec{S}_i \prod_{<k,l>} 
\left(1+\frac{1}{T}\,\vec{S}_k\cdot \vec{S}_l\right),
\label{part1}
\eeq
where $\prod_{<k,l>}$ is the product over nearest neighbour
vertices~\cite{DMNS81}. Expanding the product and integrating over spin
variables it can be seen that the only terms which survive are those
to which one can associate a collection of closed, self-avoiding and
non-intersecting loops living on the lattice and that the partition
function can also be written as~\cite{DMNS81}
\beq
Z(T)=\sum_{\{\cal L\}}\left( \frac{1}{T}\right)^{V(\{{\cal L}\})} 
n^{\cal N(\{{\cal L}\})}
\equiv \sum_{\{\cal L\}}Z_{\{\cal L\}}.
\label{part2}
\eeq
Here the sum is over all loop configurations $\{\cal L\}$ having the
above mentioned properties, $V(\{\cal L\})$ is the total number of
vertices visited by loops and ${\cal N}(\{{\cal L}\})$ is the number of
loops in $\{{\cal L}\}$. The representation~\rf{part2} makes it
possible to extend the definition of the $O(n)$-model to non-integer
and to negative values of $n$. It is well-known that for $n\in [-2,2]$
the $O(n)$-model has a second
order phase transition at some critical point
$T=T_c(n)$~\cite{Nie82}. The model is also critical for
$T<T_c(n)$~\cite{Nie82} and in this region of the coupling constant
space it is denoted as the densely packed loop model, the name
referring to the fact that the average length of the loops diverges
while the number of vertices not visited by a loop stays finite (but
non zero). The densely packed loop model describes a conformal field
theory with central charge, $c_d$, related to $n$ in the following 
way~\cite{DF84}
\beq
c_d=1-\frac{6\nu^2}{1-\nu},\hspace{0.7cm} n=2\cos(\nu \pi), 
\hspace{0.5cm} \nu\in [0,1].
\label{cd}
\eeq
If we set $T=0$ in~\rf{part1} or~\rf{part2} we obtain the fully
packed loop model. In this case only configurations where {\it all}
vertices are visited by a loop contribute to the partition function.
The fully packed loop model has been shown to describe a conformal
field theory with central charge, $c_f$ given by~\cite{BN94,BSY94}
\beq
c_f=c_d+1=2-\frac{6\nu^2}{1-\nu},
\hspace{0.7cm} n=2\cos (\nu \pi),\hspace{0.5cm} \nu\in [0,1].
\label{cf}
\eeq
If we send $n$ to zero in the fully packed loop model only
configurations with one single loop visiting all vertices exactly once
survive. In this case our partition function thus counts the number of
Hamiltonian cycles on the honeycomb lattice. According to~\rf{cf}
this statistical mechanical model has  central charge $c_f=-1$.
This $n\to 0$ limit corresponds in practice to keeping the linear 
term in $n$ of the partition function~\rf{part2}. 
For $n$ {\it identically equal} to zero, one has $c_f=c_d=0$.

\begin{figure}[htb]
\centerline{\epsfxsize=14.truecm\epsfbox{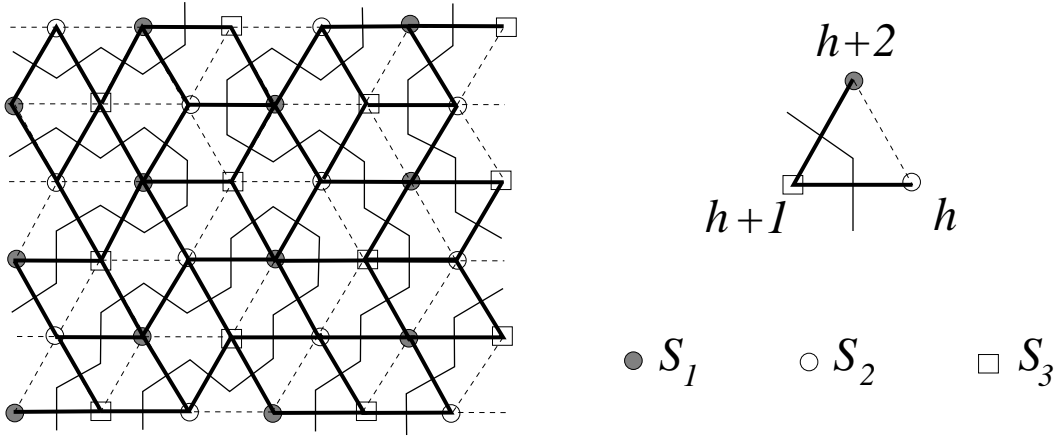}}
\caption{Construction of the SOS degree of freedom from the fully packed
loops, on a regular triangular lattice. The thick lines indicate links 
whose dual link is occupied by a loop. The height variable $h$, defined 
on the vertices of the lattice, is such that the difference of height is 
$\pm 1$ for nearest neighbours connected by a thick line and $\pm 2$ for 
nearest neighbours connected by a dashed line. The sign is fixed by 
demanding that $h$ and the number $i$ of the corresponding sub-lattice 
$S_i$ are equal modulo 3. Viewing the thick line configurations as a 
3D-piling of elementary cubes, the height variable corresponds precisely 
to the height of the surface of the piling.}
\label{SOS}
\end{figure}

An explanation of the difference in central charge between the
densely packed and the fully packed $O(n)$ model on the honeycomb
lattice was given by Bl\"ote and Nienhuis~\cite{BN94}. The fully
packed loop model can be viewed as a superposition of a densely packed loop
model and a SOS-model with central charge $c=1$. The extra SOS degree
of freedom is a height variable which emerges only in the case of
full packing. This height variable, $h_v$, lives on the vertices $v$
of the dual triangular lattice. It takes integer values and 
is defined by the demand that for 
nearest neighbour vertices $v$ and $v'$, $|h_v-h_{v'}|=1$ if the link
between $v$ and $v'$ has its dual link occupied by a loop and
$|h_v-h_{v'}|=2$ if not~\cite{BN94}. These local rules fix the value of
the height variable without ambiguity on the entire lattice, up to a global 
additive constant and a global reversal of all the signs of
$(h_v-h_{v'})$. We note that the assignment of height variables divides 
the vertices of the triangular lattice into three different sets in such 
a way that no two neighbouring vertices belong to the same set: 
if we denote the three sets as $S_i$, $i=1,2,3$ the vertices belonging 
to $S_i$ can be characterised by having $h_v=i$ (mod 3).
Therefore, the possibility of constructing the SOS height variable without 
frustrations essentially relies on the fact that the triangular lattice 
is vertex-three-colourable, i.e. can be divided into three sub-lattices of 
different colour with any two adjacent sites in different sub-lattices.
The SOS model defined here is equivalent to the zero-temperature 
anti-ferromagnetic Ising model on the triangular lattice~\cite{BH82} 
and has central charge $c=1$~\cite{BN93}. A nice way to visualise
the height variables $h_v$ is to mark only those links of the triangular
lattice which have $|h_v-h_{v'}|=1$. This leads to a picture of
a 3D-piling of cubes, whose surface profile is precisely described by 
the heights $h_v$, see figure~\ref{SOS}.

\newsection{Dense versus full packing of the $O(n)$-model on a random
lattice \label{random} }

By duality we can view the $O(n)$-model on the honeycomb lattice as
being defined on a regular triangulation and with that interpretation
the model can be coupled to quantum gravity by the standard
recipe. This leads to the so-called $O(n)$-model on a random lattice
whose partition function reads~\cite{Kos89}
\beq
Z(\lambda,T)=\sum_{\tau\sim S^2}e^{-\lambda N_{\tau}}\sum_{\{{\cal L}\}}
\frac{1}{C_{\tau}(\{{\cal L}\})} Z_{\{{\cal L}\}}.
\label{randompart1}
\eeq
Here we sum over all triangulations, $\tau$ with the topology of the sphere $S^2$
and over all loop configurations $\{{\cal L}\}$ where it is understood
that the loops live on the lattice dual to the triangulation. The
quantity $C_{\tau}(\{{\cal L}\})$ is the order of the automorphism
group of the triangulation $\tau$ with the loop configuration 
$\{{\cal L}\}$, $N_{\tau}$ is the number of triangles in the
triangulation and $\lambda$ is the cosmological constant. Let us denote
the triangles traversed by loops as decorated triangles and those not
traversed by loops as non-decorated triangles. Then we can also write
the partition function as
\beq
Z(\lambda,\kappa)=\sum_{\tau\sim S^2} e^{-\lambda N_{nd}} \sum_{\{{\cal L}\}}
\frac{1}{C_{\tau}(\{{\cal L}\})}\, \kappa^{N_d}\, n^{{\cal N}(\{{\cal L}\})},
\hspace{0.7cm} \kappa=e^{-\lambda}\,\frac{1}{T},
\label{randompart2}
\eeq
where $N_{nd}$ is the number of non-decorated triangles and $N_d$ is
the number of decorated triangles. In the coupling constant space 
($\lambda,\kappa$) of
the model~\rf{randompart2} there is a line of critical points beyond
which the partition function does not exist. On this critical line
there is a particular point $(\lambda^*,\kappa^*)$ where a phase transition
takes place. This phase transition is the analogue of the phase
transition at $T=T_c(n)$ seen on a regular lattice. For
$\kappa<\kappa^*$ the singular behaviour of the partition function is
due to the radius of convergence in $\lambda$ being reached while for 
$\kappa>\kappa^*$ the singular behaviour of the partition function is
due to the radius of convergence in $\kappa$ being reached. If we
approach the critical line from the region where $\kappa>\kappa^*$ 
we reach the densely packed
loop model on a random lattice. This model has a scaling behaviour
characteristic of a conformal matter field with central charge $c=c_d$
coupled to quantum gravity~\cite{Kos89,DK89}.

If we set $e^{-\lambda}$ equal to zero while keeping $\kappa$ finite we
obtain the fully packed loop model on a random lattice. This model
has the {\it same} scaling behaviour as the densely packed model on
a random lattice~\cite{Kos89}. In particular, if we send $n\rightarrow
0$ in the fully packed model we get a model describing Hamiltonian
cycles on a random three-coordinate lattice and this model has
$c=-2$~\cite{DK90,EGK98,Hig98}.

The fact that one does not see any change in the central charge when
going from the densely packed to the fully packed $O(n)$ model on a
random lattice is not in contradiction with the above mentioned
argument of Bl\"ote and Nienhuis. As we already noticed, in order for 
the construction of the extra SOS degree of freedom to work it is necessary 
that the vertices of the triangular lattice can be divided into three sets 
so that no two neighbouring vertices belong to the same set. 
A necessary condition for this requirement to be fulfilled is
that any vertex of the triangular lattice is shared by an {\it even} number 
of triangles. On a random lattice a vertex can be shared by any number
of triangles so the condition is not met, thus preventing the construction
of the SOS height variable. 

\begin{figure}[htb]
\centerline{\epsfxsize=6.truecm\epsfbox{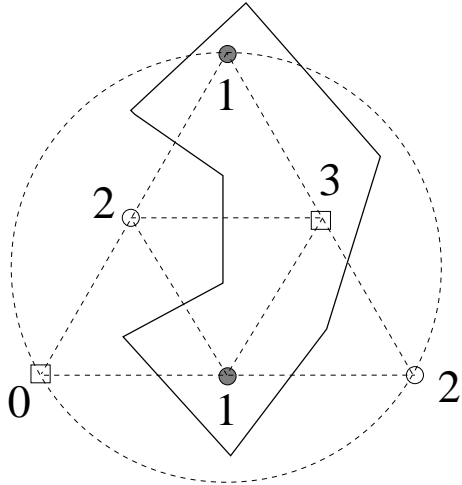}}
\caption{An Eulerian triangulation made of 8 triangles. Each vertex
is shared by an even number of triangles. We have indicated the three subsets
of different colours, as well as the height variables for 
a particular choice of Hamiltonian cycle (solid line).}
\label{EUL}
\end{figure}

The above requirement of vertex-three-colourability defines a particular
class of triangulations which we shall call Eulerian triangulations. 
Such triangulations have been studied in~\cite{DFEG98}, in connection
with the problem of folding of random lattices. For triangulations
with spherical topology, the three-colourability requirement and 
the constraint that any vertex is shared by an even number of triangles
are actually equivalent~\cite{MI97}. 
The latter condition can be rephrased by demanding
that the number of edges leaving any vertex is even, which is the well known
property ensuring that the triangulation can be drawn in one path without
lifting the pen, i.e. can be equipped with a Eulerian cycle passing
each link once.  An example of Eulerian triangulation is shown in 
figure~\ref{EUL}. 

If the argument of Bl\"ote and Nienhuis is indeed correct we should be able
to recover the shift by one in the central charge by {\it restricting}
the fully packed model to Eulerian triangulations. Indeed, for  
an Eulerian triangulation of {\it spherical topology} equipped with a fully
packed loop configuration, the construction of the SOS variable can be 
performed {\it globally} without frustration. In the case $n\rightarrow 0$ 
there is a particularly simple way of implementing this restriction to
Eulerian triangulations, as discussed below.

\newsection{Hamiltonian cycles on a random Eulerian 
triangulation \label{Eulerian} }

Let us consider the fully packed $O(n)$ model on a random Eulerian
triangulation 
in the limit $n\rightarrow 0$.  
Its partition function reads (cf.\ equation~\rf{randompart2})
\beq
Z(\kappa)=\sum_N\left(\sum_{(\tau_E\sim S^2(2N),{\cal H})}
\frac{1}{C_{\tau_E}({\cal H})}\right) \kappa^{2N}\equiv 
\sum_N Z_N\, \kappa^{2N},
\label{eulcycles}
\eeq
where the second sum is over random Eulerian triangulations of
spherical topology, equipped with a Hamiltonian cycle, ${\cal H}$,
and consisting of $2N$ triangles\footnote{Note that not all Eulerian
triangulations can be equipped with a Hamiltonian cycle.}. 
Viewing these objects as two-dimensional quantum
space-times decorated by configurations of some matter field,
it follows from the work of David, Distler and Kawai~\cite{DDK} that
$Z_N$ behaves as
\beq
Z_N\sim e^{\mu N} N^{\gamma-3} \left\{1+{\cal O}(\frac{1}{N})\right\},
\label{ddk}
\eeq
where $\gamma$ is given by
\beq
\gamma=\frac{c-1-\sqrt{(25-c)(1-c)}}{12},
\eeq
with $c$ being the central charge of the matter field. In the case of
ordinary random triangulations equipped with Hamiltonian cycles one
finds\footnote{In this case the formula~\rf{ddk} has logarithmic corrections.}  
that the exponent $\gamma$ takes the value
$\gamma=-1$~\cite{DK90,EGK98,Hig98} which is characteristic of a
matter field with $c=-2$.
What we would like to show is that in the
Eulerian case $\gamma$ takes the value characteristic of a
matter field with $c=-1$, namely the {\it irrational} value
\beq
\gamma(c=-1)=\frac{-1-\sqrt{13}}{6}.
\eeq
In order to do that we need to determine $Z_N$ in~\rf{eulcycles}. 
Therefore, let us consider a random Eulerian triangulation of spherical 
topology equipped with a Hamiltonian cycle and consisting of $2N$ triangles. 
Since the triangulation is Eulerian, we can introduce a
three-colouring of its vertices and this three-colouring of the vertices  
assigns to each triangle one of two possible orientations, $+$ and $-$. A
triangle is said to have orientation $+$ if one encounters the colours
of its vertices in the cyclic order 1-2-3 when moving counterclockwise
along its edges. If the colours are encountered in the cyclic order 1-3-2 the
triangle is said to have orientation $-$. Obviously, neighbouring
triangles have opposite orientations and any Hamiltonian cycle drawn
on the Eulerian triangulation visits alternatively $+$ and $-$ triangles.

\begin{figure}
\centerline{\epsfxsize=10.truecm\epsfbox{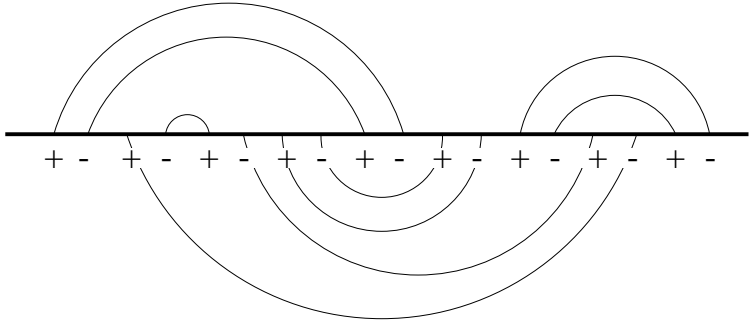}}
\caption{An arch configuration representing a planar Eulerian 
triangulation consisting of 18 triangles and being equipped with a
Hamiltonian cycle. The $+$ and $-$ signs alternate along the straight
line. Each arch connects a $+$ to a $-$.} 
\label{arches}
\end{figure}

We can now use a particular representation which facilitates our analysis. 
Let us represent the Hamiltonian cycle by a straight line having $2N$
vertices of alternating signs. The vertices represent
the triangles visited by the loop and the signs the orientation of
these triangles. As mentioned above, neighbouring triangles
necessarily have opposite orientation. So far each triangle has only
been given
two neighbouring triangles. To completely specify a triangulation we
must associate to each triangle one more neighbour. This can be done
by connecting the vertices pairwise by arches. To ensure the planarity
of the triangulation, vertices can only be connected either above or
below the straight line representing the Hamiltonian cycle and
different arches cannot intersect. Furthermore, an arch must
always connect a $+$ to a $-$. This results in arch configurations of
the type shown in figure~\ref{arches}. 
Counting such arch configurations is equivalent to performing the second
sum in equation~\rf{eulcycles}. The symmetry factor is automatically taken
into account. However,
two arch configurations differing only by
a cyclic permutation of vertices or by a reflection in the straight
line represent the same triangulation. Hence, if we denote the total
number of arch configurations involving $2N$ vertices as $M_N^E$ we have
\beq
M_N^E=2\times 2N\times Z_N\sim e^{\mu_E N} N^{\gamma_E-2}
\left\{1+{\cal O}\left(\frac{1}{N}\right)\right\}.
\label{scal}
\eeq
We note that the problem of counting
Hamiltonian cycles on an ordinary random lattice can likewise be
reduced to a problem of counting arch configurations. The arch
configurations appearing in that problem, however, are simpler than
the present ones, having no signs associated with the vertices. 
The number of such
configurations, $M_N$, can easily be counted using the 
Catalan numbers, $\{c_N\}$. One finds~\cite{EGK98,Hig98}
\beq
M_N=c_N\, c_{N+1},
\label{Mn1}
\eeq
where
\beq
c_N=\frac{1}{N+1}\left(\begin{array}{c}\!2N\! \\
\!N\!\end{array}\right)
\sim 4^N\, N^{-3/2} \hspace{0.3cm} \mbox{as} \hspace{0.3cm}
N\rightarrow \infty,
\label{Mn2}
\eeq
leading to $\gamma-2=2\times(-3/2)$, hence $\gamma=-1$ as stated previously.
In the present case where we expect to find an irrational value of $\gamma$ 
it is unlikely that a simple counting procedure can be found. 
In the next section
we shall present some analytical considerations on the counting
problem and describe a numerical solution.

\newsection{Numerical solution \label{solution}}
Our starting point is a vertex configuration of the type described above, 
i.e.\ a straight line with $2N$ vertices of alternating signs. 
We want to determine $M_N^{E}$, i.e. the number of ways of connecting pluses 
to minuses using two systems of non-intersecting arches, one drawn
above the line and one drawn below the line. 
First, let us note that we have the obvious inequality $M_N^E\leq M_N$
from which we get the upper bound $\mu_E\leq \log(16)$ (cf.\
equations~\rf{Mn1} and~\rf{Mn2}). Next, let us derive a lower bound on $\mu_E$. 
For
that purpose we consider the situation where all arches are drawn,
say, above the line. In this case the requirement of non-intersection
of the arches automatically leads to a configuration where pluses are
connected to minuses since any arch must enclose an even number of
vertices and hence connects a vertex at an odd position to a vertex at
an even position. The number of such one-side non-intersecting 
arch configurations linking $2N$ points is exactly the $N$'th Catalan
number, $c_N$. {}From a given one-side arch configuration we can
construct a two-side arch configuration  by flipping a number of
arches to the other side of the line. The resulting arch configuration
obviously contributes to $M_N^E$. Since there are $2^N$ ways of
flipping $N$ arches, we immediately deduce that $M_N^E\geq 2^N c_N$
from which we obtain the lower bound $\mu_E\geq
\log(8)$. Unfortunately, the flipping construction is not exhaustive
since it produces only configurations where the upper and lower arches
do not overlap, i.e.\ configurations where any vertex below a given,
say, upper arch is connected to another vertex below the same arch
even if the connection is made with a lower arch.

We shall describe below a systematic way of counting all the allowed
configurations. First, let us note that $M_N^E$ can be written as
\beq
M_N^E=\sum_{k=0}^N C_{N,k}
\label{MNE}
\eeq
where $C_{N,k}$ is the number of two-side arch configurations
involving $2N$ vertices and having $k$, say, upper arches. We
have the obvious symmetry
\beq
C_{N,k}=C_{N,N-k}.
\eeq
It is possible to calculate $C_{N,k}$ explicitly for
small values of $k$, as illustrated in Appendix A. 
For the first values of $k$, $C_{N,k}$ read explicitly:
\bea
C_{N,0}&=&c_N,
\nonumber \\
C_{N,1}&=&N\,c_N,
\nonumber \\
C_{N,2}&=&\frac{(2N-2)!}{(N-2)!(N+2)!}\left(3N^3-2N^2+4N\right),
\label{cnks} \\
C_{N,3}&=&\frac{(2N-3)!}{(N-2)!(N+3)!}\left(\frac{11}{3} N^6-15 N^5 
+\frac{179}{3} N^4 -117 N^3 +\frac{638}{3}N^2-192 N\right)
\nonumber \\
& &-N\cdot 4^{N-2}.
\nonumber
\eea
Besides the trivial case $k=0$, the case $k=1$ can be easily understood
by noticing that a two-side arch configuration with only one arch on
top cannot have overlapping upper and lower arches. Therefore these
arch configurations are exactly the arch configurations produced by
starting from one of the $C_{N,0}=c_N$ arch configurations without arches on  
top and flipping one of its $N$ arches to the top. This simple procedure
breaks down as soon as $k>1$ and one has to recourse to a more involved
strategy described in Appendix A. As $k$ increases, however, 
one very soon runs into rather severe complications which limit
the computation of explicit formulas to the very first values of $k$. 
In any case, for any {\it finite} $k$, one obtains the large $N$ 
scaling $C_{N,k}\sim 4^N$, far below the lower bound on $\mu_E$
and we expect that the main contribution to $M_N^E$ comes from terms 
$C_{N,k}$ with $k\approx N/2$. Unfortunately, it does not seem possible 
to calculate such terms analytically. We are not discouraged by this fact, 
however, because, as mentioned earlier, an irrational value of $\gamma$ 
is unlikely to be reproduced by a simple counting argument.

\begin{figure}[htb]
\begin{center}
\mbox{
\epsffile{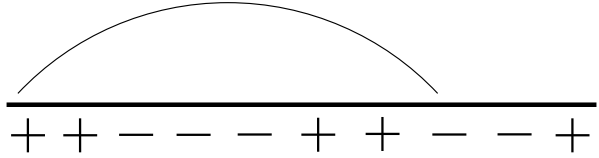}
}
\end{center}
\caption{An example of the decomposition of a sub-diagram.}
\label{example}
\end{figure}

We shall now describe how one can calculate {\it numerically} $C_{N,k}$ for
(in principle) any $N$ and $k$ using a recursive strategy. Again, let
us start from a vertex configuration having $2N$ vertices of
alternating signs and let us choose $k$ pluses and $k$ minuses
which are to be connected above the straight line. The remaining $N-k$
pluses and $N-k$ minuses are then to be connected below the straight
line. Instead of our original diagram we now have two sub-diagrams
which are to be equipped by one-side arch configurations connecting
pluses to minuses. Let us denote the number of ways of completing with 
arches a sub-diagram consisting of $k$ pluses and $k$ minuses as
$c_{\sigma_1\sigma_2\ldots \sigma_{2k}}$
where $\sigma_i$ is $+1$ or $-1$ according to whether the vertex at
position $i$ in the sub-diagram is a plus- or a minus-vertex. 
For $\sigma_i=(-1)^i$, $c_{\sigma_1\sigma_2\ldots\sigma_{2k}}$ is
nothing but the $k$'th Catalan number, $c_k$. Obviously,  
$c_{\sigma_1\sigma_2\ldots\sigma_{2k}}$ is non zero if and only if
the sequence of $\sigma_i$'s is globally neutral. There is no closed
formula giving $c_{\sigma_1\sigma_2\ldots\sigma_{2k}}$ for an arbitrary
sequence of $\sigma_i$'s. This is to be contrasted with the reverse 
problem, i.e. counting the number of $+/-$ sequences compatible 
with a given arch configuration, which is simply $2^N$ for $N$
arches since each arch must connect a plus at its left extremity to 
a minus at its right extremity or conversely. This implies in
particular the following sum rule
\beq
\sum_{\{\sigma\}}c_{\sigma_1\sigma_2\ldots\sigma_{2N}}=
\sum_{\rm arch\ configurations} 2^N\,  = \, 2^N\,c_N.
\label{sumrule}
\eeq
Still, the $c_{\sigma_1\sigma_2\ldots\sigma_{2k}}$
fulfill a certain recursion relation which can easily be implemented in
a computer program. To derive this recursion relation, let us consider
a typical sub-diagram, such as the one depicted in
figure~\ref{example}. 
In this diagram the first plus can be connected
to any minus at an even position (if the plus is connected to a minus
at an odd position it is impossible to complete the diagram without
crossings). Placing the first arch divides the remaining vertices into
two groups, those inside the arch and those to the right of the arch.
These two groups of vertices can be considered separately since there
can be no arches connecting them. This means that the
$c_{\sigma_1\sigma_2\ldots \sigma_{2N}}$ 
fulfill the following recursion relation
\begin{equation}
c_{\sigma_1\sigma_2\ldots\sigma_{2N}} =
\sum_{j=1}^{N} \delta_{\sigma_1+\sigma_{2j},0}\cdot
c_{\sigma_2\ldots\sigma_{2j-1}}\cdot c_{\sigma_{2j+1}\ldots\sigma_{2N}},
\label{recursion}
\end{equation}
where we have used the convention that $c_{\O}=1$. This recursion
relation generalises the well known quadratic relation for Catalan numbers.
To calculate $M_N^E$ numerically we now use the following obvious strategy
\begin{itemize}
\item Let $k$ run from $0$ to $\lfloor\frac{N}{2}\rfloor$, and for
each $k$, select in all possible ways $k$ pluses and $k$ minuses to be
connected above.
\item Using the recursion relation compute the numbers of ways of 
completing the upper and
lower sub-diagram that arise, and multiply these two numbers.
\item Sum the results for all choices of sub-diagrams and all values
of $k$.
\item Multiply by 2. If $N$ is even,
subtract the $k=\frac{N}{2}$-term once.
\end{itemize}
In practice, to make use of the recursion formula~\rf{recursion} we 
convert the string $\sigma_1\sigma_2\ldots\sigma_{2N}$
into a number ${\cal N}$ via the formula:
\begin{equation}
{\cal N} = \sum_{i=1}^{2N} \left(\frac{\sigma_i+1}{2}\right)\cdot 2^{i-1},
\end{equation}
and then save the number of ways of connecting the sub-diagram in a
array under position ${\cal N}$. Our program pre-computes the
number of connections for sub-diagrams up till ${\cal N}=2^{20}$ in order to 
save 
computer-time. Then formula (\ref{recursion}) is only used, if
a sub-diagram has a higher ${\cal N}$ than this, and such a sub-diagram is
quickly reduced to diagrams having ${\cal N}$ less than $2^{20}$, where the
pre-computed values can be used.

\vspace{15pt}
\noindent

\begin{figure}[htb]
\centerline{\epsfxsize=10.truecm\epsfbox{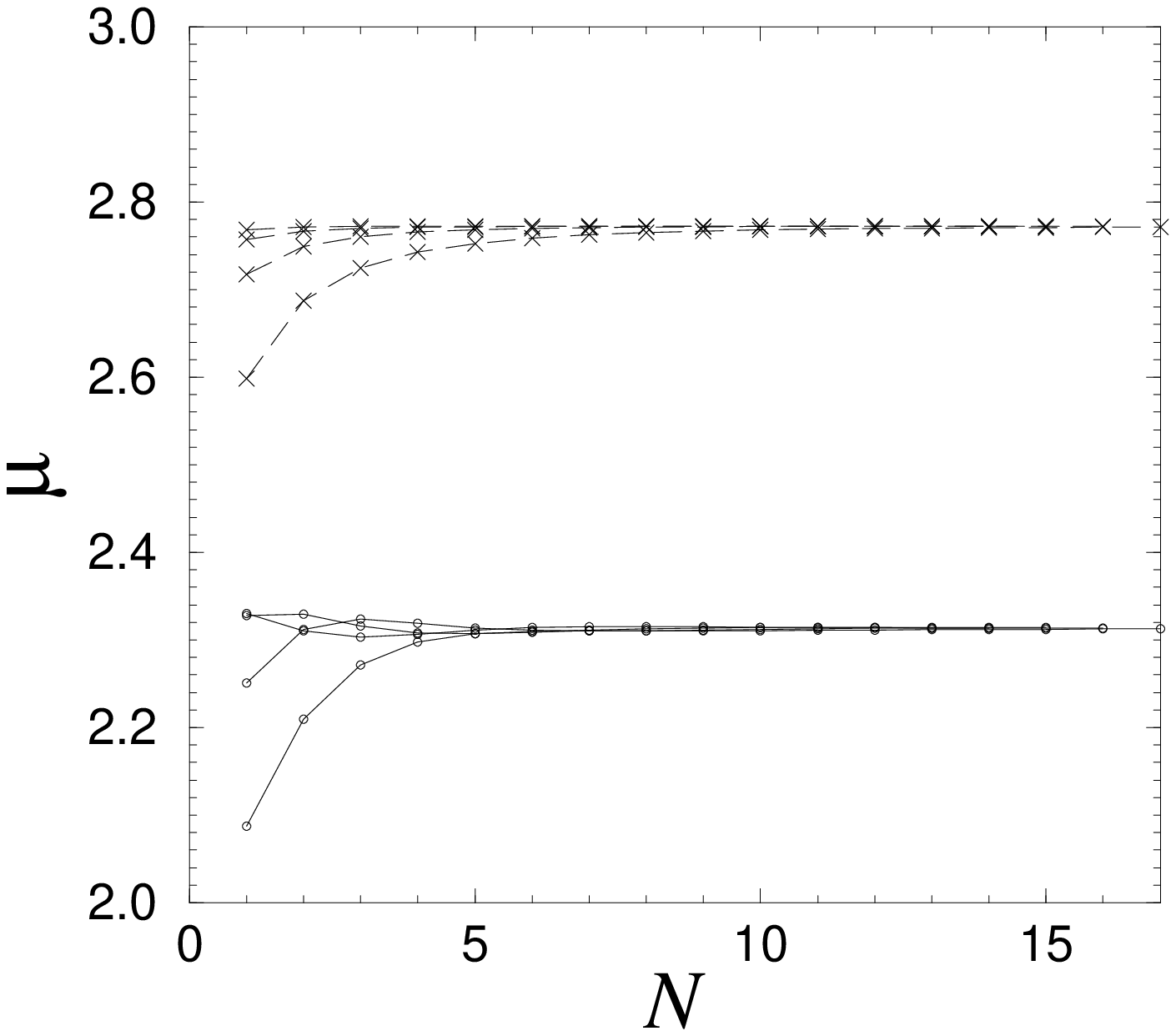}}
\caption{$\mu_N^{(p)}$ as a function of $N$ for $p=2$, 3, 4, 5 for
Hamiltonian cycles on ordinary random triangulations (dashed lines) 
and on random Eulerian
triangulations (full lines). }
\label{resmu}
\end{figure}

\begin{figure}[htb]
\centerline{\epsfxsize=10.truecm\epsfbox{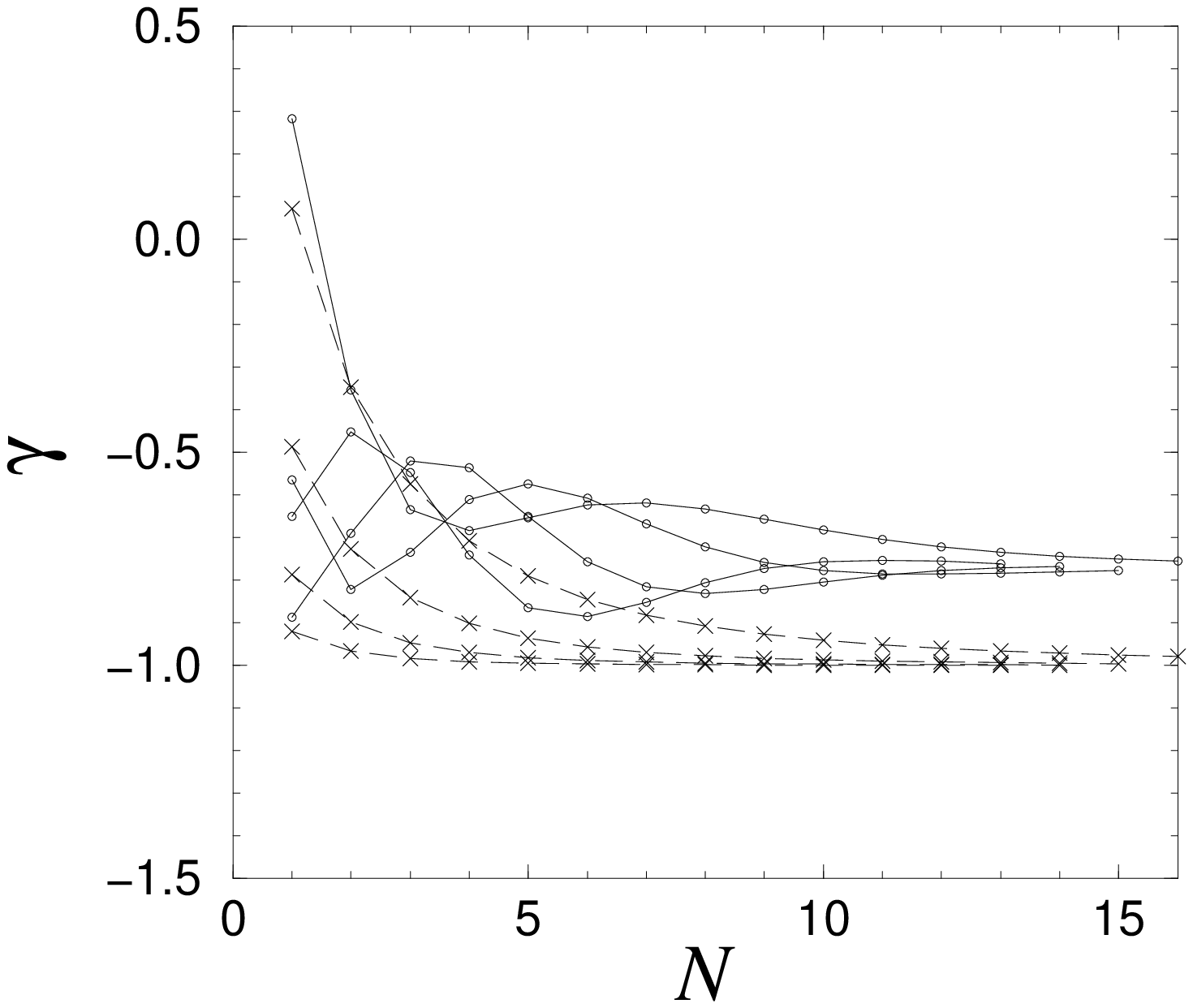}}
\caption{$\gamma_N^{(p)}$ as a function of $N$ for $p=2$, 3, 4, 5 for
Hamiltonian cycles on ordinary random triangulations (dashed lines) 
and on random Eulerian
triangulations (full lines).  }
\label{resgamma}
\end{figure}

\newsection{Results \label{results} }
In table~\ref{tabresult} we list $M_N^E$ as a function of $N$ for
$N=1,2,\ldots,20$ as determined by the computer algorithm  described in the
previous section. For comparison we list also the corresponding
number, $M_N$, counting arch systems describing ordinary random triangulations 
equipped with Hamiltonian cycles (cf.\ equation~\rf{Mn1}). Furthermore, in
table~\ref{tabresult2} we list $C_{N,k}$ for $N\leq 12$ (cf.\
equation~\rf{MNE}).
We note that
in accordance with our expectations
(cf.\ section~\ref{solution}) we observe that the main contribution to
$M_N^E$ comes from the $C_{N,k}$'s with $k\approx \frac{N}{2}$.
{}From the data for $M_N^E$ we shall seek to extract the quantities $\mu_E$ and
 $\gamma_E$ (cf.\ equation~\rf{scal}) using a variant of a ratio method
introduced in the study of random surface models in~\cite{num} (for a
general discussion, see~\cite{Sha55}). 
For a generic $M_N\sim e^{\mu N} N^{\gamma-2}
\{1+{\cal O}(1/N)\}$, we have that
\bea
\mu_N^{(0)}&\equiv &\log\left(\frac{M_{N+1}}{M_N}\right)=
\mu + {\cal O}\left(\frac{1}{N}\right),
\\
\gamma_N^{(0)}& \equiv & 2-N^2\log \left(\frac{M_{N+2} M_N}{M_{N+1}^2} \right) 
= \gamma + {\cal O}\left(\frac{1}{N}\right).
\eea
{}From these series of estimates we can get improved series of
estimates, for instance the series $\gamma_N^{(p)}$ given by
\beq
\gamma_N^{(p)}\equiv \frac{1}{p!}\sum_{i=0}^p (N+i)^p 
\left(\begin{array}{c}\!p\! \\ \!i\!\end{array}\right) \, (-1)^{p-i} 
\gamma_{N+i}^{(0)}=\gamma +{\cal O}(\frac{1}{N^{p+1}}),
\eeq
and similarly for $\mu$.
{\footnotesize
\begin{table}
\begin{tabular}{|r|r|r|} \hline
 & & \\
$N$ & $M_N^E$ & $M_N$ \\
& &
\\ \hline \hline
1 & 2 & 2\\ \hline 
2 & 8 & 10 \\ \hline
3 & 40 & 70 \\ \hline
4 & 228 & 588 \\ \hline
5 & 1424 & 5544 \\ \hline
6 & 9520 & 56628 \\ \hline
7 & 67064 & 613470\\ \hline
8 & 492292 & 6952660  \\ \hline
9 & 3735112 & 81662152\\ \hline
10 & 29114128 & 987369656\\ \hline
11 & 232077344 & 12228193432\\ \hline
12 & 1885195276 & 154532114800\\ \hline
13 & 15562235264 & 1986841476000 \\ \hline
14 & 130263211680 & 25928281261800\\ \hline
15 & 1103650297320 & 342787130211150 \\ \hline
16 & 9450760284100 & 4583937702039300\\ \hline
17 & 81696139565864 & 61923368957373000\\ \hline
18 & 712188311673280 & 844113292629453000\\ \hline
19 & 6255662512111248 & 11600528392993339800\\ \hline
20 & 55324571848957688 & 160599522947154548400\\ \hline
\end{tabular}
\caption{The numbers $M_N^E$ (random Eulerian triangulations) and
$M_N$ (ordinary random triangulations) for $N=1,$ 2, $\ldots, 20$. }
\label{tabresult}
\end{table}}
{\footnotesize
\begin{table}
\begin{center}
\begin{tabular}{|c|rrrrrrrrrrrr|}
\hline
$\scriptstyle k\backslash N$& \multicolumn{1}{c}{1} & \multicolumn{1}{c}{2} & 
\multicolumn{1}{c}{3} & \multicolumn{1}{c}{4} & \multicolumn{1}{c}{5} & 
\multicolumn{1}{c}{6} & \multicolumn{1}{c}{7} & \multicolumn{1}{c}{8} & 
\multicolumn{1}{c}{9} & \multicolumn{1}{c}{10} & 
\multicolumn{1}{c}{11} & \multicolumn{1}{c|}{12} \\ \hline
0 & 1 & 2 & 5  & 14 & 42  & 132  & 429   & 1430   & 
4862    & 16796   & 58786    & 208012    \\ 
1 & 1 & 4 & 15 & 56 & 210 & 792  & 3003  & 11440  & 
43758   & 167960  & 646646   & 2496144   \\
2 &   & 2 & 15 & 88 & 460 & 2250 & 10549 & 48048  & 
214344  & 941460  & 4085950  & 17566032  \\
3 &   &   &  5 & 56 & 460 & 3172 & 19551 & 111584 & 
602514  & 3121020 & 15655970 & 76559920  \\
4 &   &   &    & 14 & 210 & 2250 & 19551 & 147288 & 
1002078 & 6320460 & 37614016 & 213817902 \\
5 &   &   &    &    &  42 & 792  & 10549 & 111584 & 
1002078 & 7978736 & 57977304 & 392238792 \\
6 &   &   &    &    &     & 132  & 3003  & 48048  &
602514  & 6320460 & 57977304 & 479421672 \\
7 &   &   &    &    &     &      & 429   & 11440  & 
214344  & 3121020 & 37614016 & 392238792 \\
8 &   &   &    &    &     &      &       &  1430  & 
43578   & 941460  & 15655970 & 213817902 \\
9 &   &   &    &    &     &      &       &        & 
4862    & 167960  & 4085950  & 76559920  \\
10 &  &   &    &    &     &      &       &        & 
        & 16796   & 646646   & 17566032  \\
11 &  &   &    &    &     &      &       &        & 
        &         & 58786    & 2496144   \\
12 &  &   &    &    &     &      &       &        &     
        &         &          & 208012    \\ \hline
\end{tabular}
\end{center}
\caption{The numbers $C_{N,k}$ for $N\leq12$. }
\label{tabresult2}
\end{table}}
\noindent
In figures~\ref{resmu} and \ref{resgamma}, we plot 
$\mu_N^{(p)}$ and $\gamma_N^{(p)}$ as a
function of $N$ for $p=2,3,4,5$ for the Hamiltonian cycles
on the ordinary random triangulations (dashed lines) as well as for
the Hamiltonian cycles on the random Eulerian triangulations (full lines).
The series $\mu_N^{(p)}$ and $\gamma_N^{(p)}$ are expected to converge
faster the larger the value of $p$. However, the larger $p$ the
 smaller the amount of data. 

In table~\ref{final} we list our final estimates for $\mu$ and
$\gamma$ extracted from the series $\mu_N^{(p)}$ and $\gamma_N^{(p)}$
for ordinary random triangulations (RT) and random Eulerian 
triangulations (RET). We also give for comparison the exact result for
ordinary random triangulations.

\begin{table}
\vspace*{1,0cm}
\begin{tabular}{|l|c|c|}
\hline
& $\mu$ & $\gamma$  \\ \hline 
 RT (exact) & $\log(16)\approx 2.772589$ & $-1$  \\ \hline
 RT (estimated) &   $ 2.772590 \pm 10^{-6}$ & $-1.000000 \pm 10^{-6}$ \\ \hline
 RET        & $2.313\pm 0.001$ & $-0.76 \pm 0.02$ \\ \hline
\end{tabular}
\caption{Final estimates for $\mu$ and $\gamma$ for Hamiltonian cycles
on ordinary random triangulations (RT) and random Eulerian
triangulation (RET) along with the exact results for the RT-case.}
\label{final}
\end{table}
The ratio method reproduces neatly the exact values for the ordinary
random triangulations. When we make the restriction to random Eulerian
triangulations we clearly see a difference in the behaviour of
$\mu_N^{(p)}$ and $\gamma_N^{(p)}$. In the case of $\mu_N^{(p)}$ the
convergence is as rapid as for ordinary random triangulations but the
asymptotic value is smaller. We note that our estimate $\mu_E\sim 2.313$ 
respects the bounds $\log(8)\leq \mu_E\leq \log(16)$ derived in 
section~\ref{solution}, as it should. In
the case of $\gamma_N^{(p)}$ we observe an oscillatory behaviour, not
present for the usual random triangulations. Since the
amplitude of the oscillations decreases as $N$ increases the data
still allows us to extract a reliable estimate for $\gamma$. This
estimate for $\gamma$ agrees well with the predicted irrational value
$\gamma_E=(-1-\sqrt{13})/6\approx -0.768$ and definitely differs from
the value $\gamma=-1$ characteristic of Hamiltonian cycles on usual
random triangulations.

\newsection{Conclusion \label{conclusion} }

We have argued that one should see an increase of the central charge
by one unit when one considers the fully packed $O(n)$ model on a
random {\it Eulerian} triangulation instead of an ordinary random
triangulation. Our argument was based on an explanation of Bl\"{o}te
and Nienhuis why one sees an increase in central charge when moving
from the densely packed to the fully packed $O(n)$ model on a regular
lattice~\cite{BN94} (cf.\ section~\ref{regular}). We have tested our
prediction by numerical analysis in the case $n\rightarrow 0$ which 
corresponds to Hamiltonian cycles on random Eulerian triangulations
and on ordinary random triangulations respectively. Whereas the latter
system has $c=-2$ the first one should according to our prediction
have $c=-1$. Our numerical analysis confirms the prediction. In
particular this means that we have found an explicit realisation of
a random surface model with an {\it irrational} value of the entropy
exponent, $\gamma$, not at any point evoking analytical continuation. 
In addition, our results support the explanation 
of Bl\"ote and Nienhuis. 

Let us denote by ${\cal M}_N(q)$ the number of Hamiltonian cycles on a
lattice with $N$ sites and coordination number $q$. A mean field
argument of H.\ Orland et al.~\cite{OID85} gives that
\beq
\omega_H=\lim_{N\rightarrow \infty} \frac{1}{N}
\log {\cal M}_N(q)=\frac{q}{e}.
\eeq
In the present case we find that the number of Hamiltonian cycles on
random Eulerian triangulations behaves as
$\left[\exp(2.313)\right]^{T/2}$
where $T$ is the number of triangles or equivalently the number of
vertices on the dual lattice. The number of random Eulerian
triangulations consisting of $T$ triangles, on the other hand, is
known to behave as $\sqrt{8}^T$ \cite{DFEG98}. This gives the following
average value of $\omega_H$
\beq
\langle \omega_H \rangle \approx
\left(\frac{e^{2.313}}{8}\right)^{1/2}=1.124,
\eeq
which is very close to the mean field value $\frac{3}{e}\approx 1.104$.
For the honeycomb lattice one has $\omega_H=\frac{3^{3/4}}{2}\approx
1.140$~\cite{BSY94}.

It would be interesting to extend our analysis of the $O(n)$ model
on random Eulerian triangulations to other values of $n$.
The case $n=2$ is particularly interesting since, as explained in the
introduction, this case describes the folding problem of a random
triangulation onto itself, modelling a fluid membrane. 
As likewise explained in the introduction, this folding
problem can be viewed as an edge-three-colouring problem on a random
vertex-three-colourable triangulation. The system of
vertex-three-colourable triangulations has been found to have central
charge $c=0$~\cite{DFEG98} and so has the edge three-colouring problem 
on a usual random triangulation~\cite{EK97}. Amazingly, when coupling
these two models one should get a model with central charge $c=2$.

\begin{figure}[htb]
\centerline{\epsfxsize=8.truecm\epsfbox{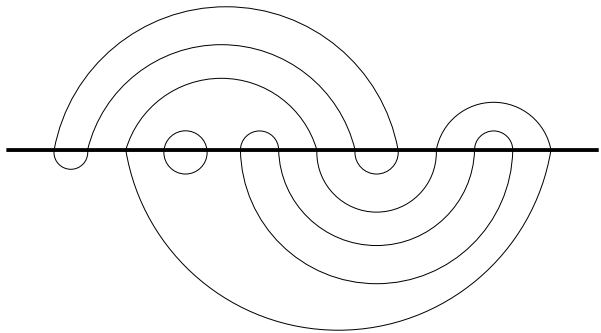}}
\caption{ An example of a Meander configuration with 
$4$ connected components.}
\label{meandre}
\end{figure}

Finally, we would like to make a connection between our problem
and another problem of arches, that of Meanders.  Originally, 
the Meander problem consists in enumerating the number of pairs of 
arch systems connecting $2N$ points {\it both} from above and from 
below and such that the path resulting from connecting the upper
and lower arches is made of a single connected component. 
This problem turns out to be equivalent to the compact folding 
{\it onto itself} of a closed linear self-avoiding polymer. 
By extension, one also considers the possibility of creating 
several connected components, with a weight $q$ per connected component. 
Figure~\ref{meandre} shows an example of a Meander configuration with $4$ 
connected 
components.  The partition function one wants to evaluate is therefore
\begin{equation}
z_N(q)=\sum_{a,b}q^{c(a,b)},
\label{meander}
\end{equation}
where $a$ and $b$ are arbitrary systems of arches connecting $2N$ points
on one side (say the upper side for $a$ and the lower side for $b$)
of a straight line, and $c(a,b)$ is the number of connected components
obtained by connecting these arches. As for our problem, one expects
$z_N(q)\sim e^{\mu(q)N}N^{\gamma(q)-2}$ at large $N$. Except for the 
particular cases $q=1$, $q=-1$ and $q\to \infty$ which can be solved explicitly,
only numerical estimates for $\mu$ and $\gamma$ are known. 
As for our system, the Meander problem is a problem of {\it interacting}
arch systems. In our case, the interaction comes from the fact that
the upper and lower arches have to connect pluses and minuses for
two sub-sequences of the {\it same} alternating sequence. In the Meander
problem, the interaction concerns the number of created connected components.
It is interesting to note that this interaction can also be reformulated
in terms of sequences of pluses and minuses to be connected by arches.
Indeed, as shown in~\cite{Meanders} (in a slightly different language),
the quantity $q^{c(a,b)}$ can be written as
\begin{equation}
q^{c(a,b)}=\sum_{\{\sigma\}\ a\ {\rm and}\ b\atop \ \ {\rm connectable}}
u^{\frac{1}{2}\left(P(a,\{\sigma\})+P(b,\{\sigma\})- Q(a,\{\sigma\})-
Q(b,\{\sigma\}\right)},
\label{connec}
\end{equation}
for $q=u+1/u$ and where the sum is over all sets $\{\sigma\}$ of pluses and
minuses which are such that both $a$ and $b$ connect only pluses to minuses
(we say: $a$ and $b$ connectable). Here $P(a,\{\sigma\})$ (respectively 
$Q(a,\{\sigma\})$) denote the number of arches in $a$ connecting a plus 
(respectively a minus) at the left extremity of the arch to a minus
(respectively a plus) at the right extremity. For $u=1$ for instance, 
this leads to 
\begin{equation}
z_N(2)=\sum_{a,b}\sum_{\{\sigma\}\ a\ {\rm and}\ b\atop \ \ {\rm connectable}} 1
=\sum_{\{\sigma\}}c_{\sigma_1\sigma_2\cdots\sigma_{2N}}^2,
\end{equation}
to be compared with equation~\rf{sumrule}.
We thus see here that one of the main difficulties in the Meander
problem actually consists in the determination of the coefficients
$c_{\sigma_1\cdots\sigma_{2N}}$ for arbitrary sequences.
In this sense, solving our problem would be a first step in the 
solution to the Meander problem.

\vspace{15pt}
\noindent
{\bf Acknowledgements} \hspace{0.3cm}
We thank J.\ Ambj\o rn, B.\ Durhuus and J.\ Jurkiewicz for useful
discussions and O. Golinelli for a critical reading of the manuscript.

\begin{figure}[htb]
\begin{center}
\mbox{
\epsfxsize=10cm
\epsffile{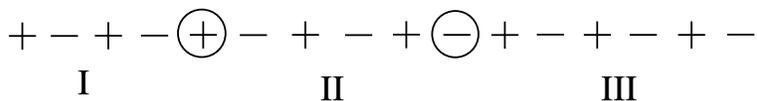}
}
\end{center}
\caption{Choosing one plus and one minus divides the
diagram into three sub-diagrams.}
\label{pm}
\end{figure}

\begin{figure}[htb]
\begin{center}
\mbox{
\epsfxsize=10cm
\epsffile{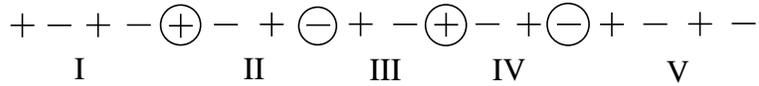}
}
\end{center}
\caption{Choosing two pluses and two minuses divides the diagram into 
five sub-diagrams.}
\label{pmpm}
\end{figure}

\appendix{}
In this Appendix, we will explain how to compute the coefficients 
$C_{N,k}$ for the first values of $k$. Let us illustrate our strategy 
by calculating $C_{N,1}$ in a way which can be generalised to larger
values of $k$. We again start from a vertex configuration involving
$2N$ vertices of alternating signs. Now we must choose a plus and a
minus which are to be connected above the line. There are two
different situations that we need to consider. Either the plus chosen
is situated to the left of the minus chosen or the plus chosen is
situated to the right of the minus chosen. Corresponding to these two
possibilities we write with an obvious notation
\beq
C_{N,1}=C_{N,1}^{+-}+C_{N,1}^{-+}.
\eeq
Let us now consider the first situation (cf.\ figure~\ref{pm}). 
By choosing the plus and the minus which are to be connected above the
line we split the vertex configuration into three sub-configurations,
I, II, and III. These three sub-configurations all contain an equal
number of pluses and minuses and we shall denote such configurations
as neutral. If we considered instead the opposite situation only
the sub-diagram in the middle would be neutral. The left-most
sub-diagram would have a plus in excess and the right-most
sub-diagram would have a minus in excess. Such diagrams will be
denoted as positively and negatively charged respectively. As we shall
see later the charge of a sub-diagram plays an important role
in the counting process. Now the vertices of the three sub-diagrams
must be connected by arches drawn below the line. A vertex of
sub-diagram II cannot be connected to a vertex of sub-diagram I or
III. However, nothing prevents vertices from sub-diagram I from
being connected to vertices of sub-diagram III and we can consider the
union of these two sub-diagrams as one sub-diagram. In general a
series of sub-diagrams can be connected if the resulting diagram is one
with alternating signs. Now, we can immediately write down an
expression for $C_{N,1}^{+-}$. Instead of calculating directly
$C_{N,k}$, however, it proves convenient to calculate first the
corresponding generating functional $C_{k}(z)=\sum_{N=0}^{\infty}
C_{N,k}\, z^N$. Denoting the number of vertices in sub-diagram I as $2i$
and the number of vertices in sub-diagram II as $2j$ we have
\bea
C_{1}^{+-}(z)&=&\sum_{N=0}^{\infty}\sum_{i=0}^{N-1}\sum_{j=0}^{N-1-i}
c_j\, c_{N-j-1}z^N =\sum_{N=0}^{\infty}\sum_{j=0}^{N-1}(N-j) c_j\,
c_{N-j-1}z ^N \nonumber \\
&=&zC(z)\frac{d}{dz}\left( zC(z)\right),
\eea
where
\beq
C(z)=\sum_{N=0}^{\infty}c_N z^N=\frac{1-\sqrt{1-4z}}{2z}.
\label{Cz}
\eeq
By similar considerations one finds for $C_{N,1}^{-+}(z)$
\beq
C_{1}^{-+}(z)=\sum_{N=0}^{\infty}\sum_{i=1}^{N-1}\sum_{j=0}^{N-1-i}c_j\,
c_{N-j-1} z^N =zC(z) \left(z\frac{d}{dz} C(z)\right),
\eeq
the difference from the previous case being the lower limit in the
second sum (which is due to the fact that in this case sub-diagram I is
positively charged). Thus, in total we get
\beq
C_{1}(z)=z\frac{d}{dz}\left(z C(z)^2\right),
\eeq
which means that
\beq
C_{N,1}=N\, c_N,
\eeq
in agreement with equation~\rf{cnks}.
For $k=1$, the above calculation is clearly not the simplest way to
reach the desired result, but it has the advantage that it can be 
generalised to the $C_{N,k}$'s with a finite $k>1$ even if, as we 
shall see, one very soon runs into a rather severe complication. 
Let us illustrate this by calculating $C_{2}^{+-+-}(z)$
In this case choosing the four vertices which are to be connected
above the line divides the vertex configuration into five
sub-diagrams I, II, III, IV and V which are all neutral (cf.\
figure~\ref{pmpm}). 
It is not possible to connect a vertex of a
sub-diagram with an odd number to a vertex of a sub-diagram with an
even number, but all other combinations are allowed. For
$C_{2}^{+-+-}(z)$ one thus has, cf.\ figure~\ref{pmpm}
\bea
\lefteqn{\hspace{-1.0cm}C_{2}^{+-+-}(z)=2\sum_{N=0}^{\infty}
\sum_{i=0}^{N-2}\sum_{j=0}^{N-2-i}\,
\sum_{k=0}^{N-2-i-j}
\,\,\sum_{l=0}^{N-2-i-j-k}\left\{ c_{j+l}\, c_k\, c_{N-2-j-k-l}\right.}
\nonumber \\
 & & \hspace{1.0cm}\left.
+c_{N-2-j-l}\,c_j\, c_l -c_j\,c_k\,c_l\, c_{N-2-j-k-l}\right\}z^N.
\label{Cn2pmpm}
\eea
The first term in the sum represents the configurations where we
allow (but not enforce) connections between region II and region IV
and forbid connections between region III and region I or V, 
The second term represents the configurations where we forbid
connections between regions II and IV, allowing connections 
between region III and the regions I and V. There is an overlap between 
the configurations which contribute to these two terms and to remove 
this overlap we must subtract exactly the third term, corresponding
to situations where the regions II, III and IV are not connected 
to any other region. In the present case it is straightforward to 
remove the overlap but for higher values of $k$ the overlapping 
becomes a severe complication. 

As long as one is able to keep track of the overlaps, however, there
are simple diagrammatic rules by means of which $C_{k}(z)$ can be
constructed. Let us assume that we are aiming at constructing 
$C_{k}^{\sigma_1\sigma_2\ldots \sigma_{2k}}(z)$ where
$\sigma_1\sigma_2\ldots\sigma_{2k}$ is a given sequence of $k$ pluses
and $k$
minuses.
Choosing the $k$ pluses and the $k$ minuses which are to be connected
above the line divides the original vertex configuration into $2k+1$
sub-configurations. To calculate $C_{k}^{\sigma_1\sigma_2\ldots
\sigma_{2k}}(z)$ we must sum over all possible ways of connecting these
sub-diagrams and if necessary subtract overlapping
configurations. Assume that we have chosen one particular way of allowing
connections between the $2k+1$ sub-diagrams. Our particular choice has arranged
the sub-diagrams into groups $G_i$ of possibly connected diagrams.
The contribution to
$C_{k}^{\sigma_1\sigma_2\ldots\sigma_{2k}}(z)$ from this particular way of
connecting the sub-diagrams can be written as
\beq
z^kc_{\sigma_1\sigma_2\ldots\sigma_{2k}}\prod_i F_{G_i}(z).  \label{rule1}
\eeq
Here $c_{\sigma_1\sigma_2\ldots\sigma_{2k}}$ counts the number of ways of
connecting the $k$ pluses and $k$ minuses above the line. If the total number
of sub-diagrams in a group $G$ is denoted as $m$ and the number of
positively charged sub-diagrams is denoted as $m_+$, it is easily shown
that $F_G(z)$ takes
the form\footnote{In any given group of connected sub-diagrams the
number of positively charged diagrams must be equal to the number of
negatively charged diagrams.}
\beq
F_G(z)=\frac{1}{(m-1)!} z^{m_+}\frac{d^{m-1}}{dz^{m-1}}
\left(z^{m-1-m_+} C(z)\right).
\label{rule2}
\eeq
For instance in the case of $C_{N,2}^{+-+-}$ one finds immediately
$$
C_{2}^{+-+-}(z)=2z^2 C(z)\left[\frac{d}{dz}(zC(z))\right]^2
+2 z^2 C(z)^2 \frac{1}{2}
\frac{d^2}{dz^2}(z^2 C(z))-2z^2 C(z)^3\frac{d}{dz}(z C(z)),
$$
which is easily seen to coincide with the expression~\rf{Cn2pmpm}.
Using the rules~\rf{rule1} and~\rf{rule2} as well as the explicit
expression for $C(z)$ (cf.\ equation~\rf{Cz}) we have been able to
determine $C_{N,2}$ and $C_{N,3}$, as given in equation~\rf{cnks}.
It is obvious that as long as we consider $k$ {\it finite}, $C_{k}(z)$ can be
expressed in terms of a finite number of terms involving
only the function $C(z)$ and its derivatives and thus 
$C_{N,k}\sim 4^N$ at large $N$.

\end{document}